\documentclass[amssymb, preprintnumbers, showpacs, showkeys,aps,prl,superscriptaddress,twocolumn,longbibliography]{revtex4-2}

\usepackage{color} 
\usepackage[hidelinks, hypertexnames=false]{hyperref}
\usepackage[all]{hypcap}
\usepackage{slashed}
\usepackage{graphicx}
\usepackage{amsmath}
\usepackage{latexsym}
\usepackage{epstopdf}
\usepackage{amsmath}
\usepackage{enumitem}
\usepackage{amssymb,amsmath}
\usepackage{multirow}
\usepackage{mathtools}
\usepackage{bbm}
\usepackage{bm}
\usepackage{amsfonts}
\usepackage{amssymb}
\usepackage{calligra}
\usepackage{calrsfs}
\usepackage{makecell}
\usepackage[normalem]{ulem}
\usepackage[USenglish,american]{babel}

\newcommand*{\figref}[2][]{%
  \hyperref[{fig:#2}]{%
    Fig~\ref*{fig:#2}%
    \ifx #1\\%
    \else
      \,(#1)%
    \fi
  }%
}

\newcommand{\myparagraph}[1]{\paragraph{#1.}}

\begin{document}

\title{Finite-temperature phases of trapped bosons in a two-dimensional quasiperiodic potential}

\author{Matteo Ciardi}
\email{matteo.ciardi@unifi.it}
\affiliation{Dipartimento di Fisica e Astronomia, Universit\`a di Firenze, I-50019, Sesto Fiorentino (FI), Italy}
\affiliation{INFN, Sezione di Firenze, I-50019, Sesto Fiorentino (FI), Italy}

\author{Tommaso Macr\`i}
\email{macri@fisica.ufrn.br}
\affiliation{Departamento de F\'isica Te\'orica e Experimental, Universidade Federal do Rio Grande do Norte, and International Institute of Physics, Natal-RN, Brazil}

\author{Fabio Cinti}
\email{fabio.cinti@unifi.it}
\affiliation{Dipartimento di Fisica e Astronomia, Universit\`a di Firenze, I-50019, Sesto Fiorentino (FI), Italy}
\affiliation{INFN, Sezione di Firenze, I-50019, Sesto Fiorentino (FI), Italy}
\affiliation{Department of Physics, University of Johannesburg, P.O. Box 524, Auckland Park 2006, South Africa}

\begin{abstract}
We study a system of 2D trapped bosons in a quasiperiodic potential via {\it ab initio} Path Integral Monte Carlo simulations, focusing on its finite temperature properties, which have not yet been explored. Alongside the superfluid, normal fluid and insulating phases, we demonstrate the existence of a Bose glass phase, which is found to be robust to thermal fluctuations, up to about half of the critical temperature of the non-interacting system. Local quantities in the trap are characterized by employing zonal estimators, allowing us to trace a phase diagram; we do so for a set of parameters within reach of current experiments with quasi-2D optical confinement.
\end{abstract}

\maketitle

\myparagraph{Introduction} Quasiperiodic potentials have attracted the interest of the scientific community in recent years. Their peculiar geometrical and physical properties \cite{lev86, sen96, jan12} promise to, or have already begun to, give contributions to topics such as topological states of matter \cite{oza19, kra12}, quantum many-body localization \cite{sch15}, and several others. Originally discovered in solid state systems \cite{she84}, quasicrystalline phases have also been realized with ultracold atoms in optical traps \cite{gui97, roa08} and in various photonics setups \cite{var13} Recent experimental work has characterized a system of 3D bosons confined by a 2D quasiperiodic potential \cite{Sbroscia2020, Viebahn2019}, showing proof of a transition from an extended to a localized state for free and interacting bosons alike. These studies opened the way for a more detailed analysis of the nature of the localized phase.

The Bose glass (BG) \cite{Fisher1989} is an insulating phase with rare superfluid puddles \cite{pol09, svistunov2015superfluid}. This leads to the absence of global superfluidity, just as in an insulating phase, accompanied by a finite compressibility, which can be related to excitations in the puddles. Such a phase cannot appear in periodic systems, where rare regions are intrinsically absent, but it is predicted to emerge in the presence of disorder \cite{PhysRevLett.111.050406}; the main example is the disordered Bose-Hubbard model \cite{soy11, het18}, where it has been proved that a direct SF-MI transition cannot take place \cite{pol09}. Quasiperiodic potentials, which present long-range order but are not translationally invariant, offer an alternative geometry, capable of birthing rare regions of local superfluidity which are deterministic and constrained by long-range order.

\begin{figure}[t!]
\includegraphics[width=\linewidth]{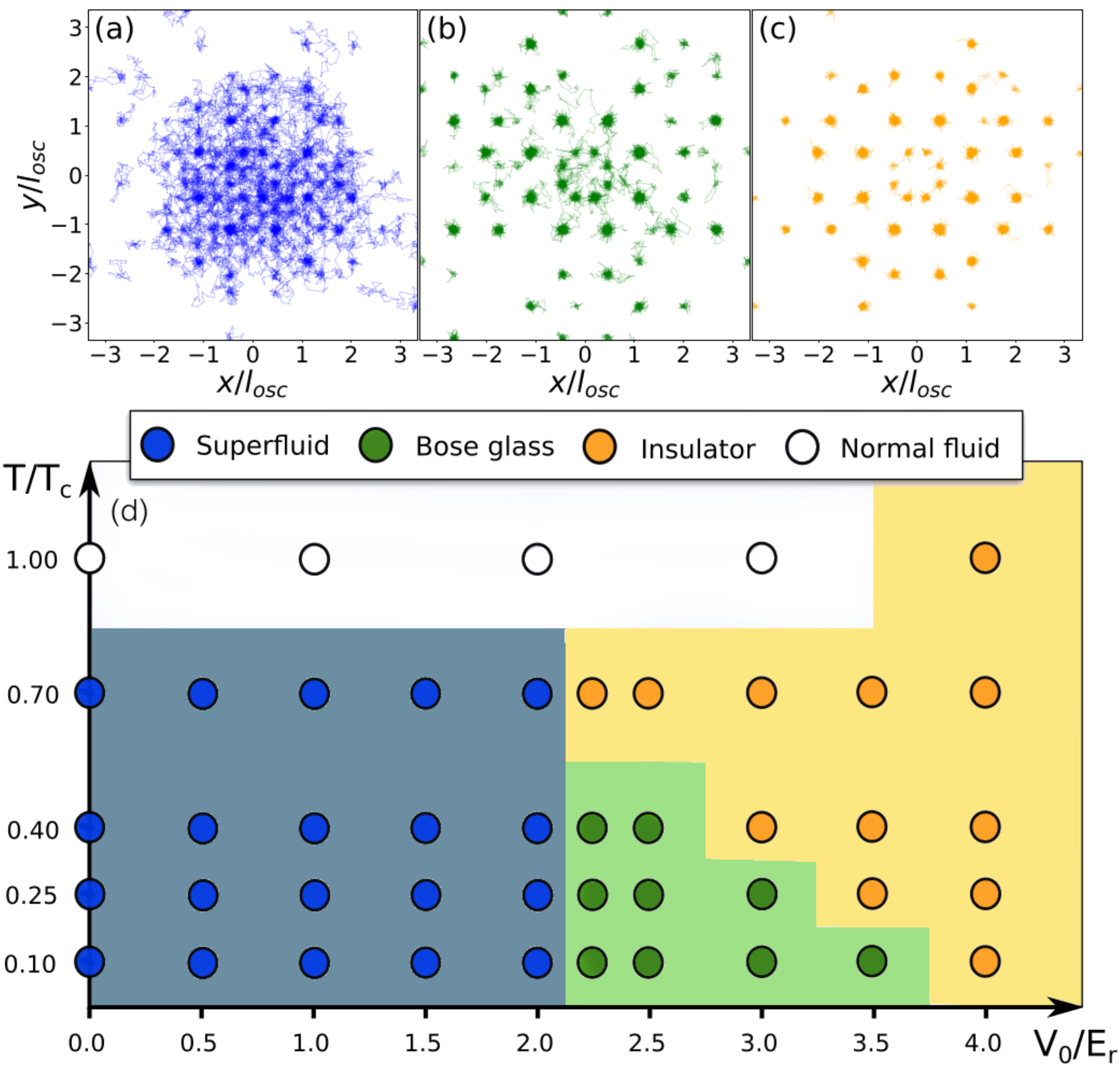}
\caption{\textit{Phase diagram.} 
(a–c): Snapshots of three configurations at $T/T_c = 0.25$, corresponding to $V_0 = 1.5 E_r$ (a), $V_0 = 2.5 E_r$ (b), and $V_0 = 3.5 E_r$ (c). Colors illustrate different phases, as discussed below.
(d): Phase diagram in the harmonic trap as a function of temperature $T$ and potential strength $V_0$. The interaction is fixed at $\tilde{g}=0.0217$. Each circle corresponds to a simulation point, and colors mark the corresponding phase: superfluid (blue), Bose glass (green), insulator (orange), or normal fluid (white). Shaded areas are a guide for the eye.}
\label{fig:phasediagram}
\end{figure}

Recently, a number of studies have investigated the properties of interacting bosons in two-dimensional quasicrystalline potentials on quasicrystalline lattices in continuous space  \cite{gau21}, in the tight-binding limit \cite{joh19}, and on a square lattice through the 2D Aubry-André model \cite{sza20, joh20}.
Some of these works have delineated phase diagrams in the mean-field approximation, and for strongly interacting particles; they have shown that regions of BG appear between the superfluid (SF) and Mott insulator (MI) phases, similarly to the disordered case. Most of these investigations have dealt with homogeneous systems, without considering the possible effects of harmonic trapping. Until now, only ground-state properties have been determined: the behavior of the system at finite temperature, which is to say the effect of thermal fluctuations on the localization transition and on the BG phase, has not yet been explored.

In this letter, we address the question of characterizing the BG phase in trapped two-dimensional systems, and we investigate its fate at finite temperature. 
% \textbf{Since the system is finite-sized, we do not speak here of phases in the sense of a thermodynamic limit; we still speak of phases, however, when they can be distinguished based on the values of our observables, since their determination is of experimental interest.}
Since the number of particles is fixed, we do not speak of phases in the sense of a thermodynamic limit. Nonetheless, their determination is of experimental interest, and we focus on parameters which could be accessed by current experiments \cite{had06, cho13, fle15}.
In this context, for a fixed value of the interaction, we trace an exact ``phase diagram'' (\figref[d]{phasediagram}), showing that a BG phase can still be identified in the presence of the harmonic trap 
%(see, e.g., \figref[b]{phasediagram} and the corresponding diffraction pattern in \figref[g]{geometry})
and that it is, up to a certain point, resilient to thermal fluctuations. 

\myparagraph{Model} We study a continuous two-dimensional model of $N$ bosons of mass $m$, subjected to an isotropic harmonic trapping of frequency $\omega$ and to an external quasiperiodic potential $V_{qc}(\mathbf{r})$. The many-body Hamiltonian reads
\begin{equation}\label{Hamiltonian}
\mathcal{H} = \sum_{i=1}^{N} \left( \frac{\textbf{p}_i^2}{2m} + \frac{m\omega^2}{2} \textbf{r}_i^2 + V_{qc}(\textbf{r}_i) \right) + \sum_{i<j} V_{int}(|\textbf{r}_i-\textbf{r}_j|),
\end{equation}
where $\textbf{r}_i$ is the position of the $i$-th particle, $\textbf{p}_i$ its momentum, and $V_{int}$ is the interaction potential between two particles. The quasiperiodic potential reads 
\begin{equation}\label{vqc}
V_{qc}(\textbf{r}) = V_0 \sum_{i=1}^{4} \cos^2 (\textbf{k}_i \cdot \textbf{r}),
\end{equation}
where the wave vectors $\mathbf{k}_i$ are given by
$ \textbf{k}_1 = k_{lat} \begin{pmatrix} 1 & 0 \end{pmatrix}$, $ \textbf{k}_2= k_{lat}/\sqrt{2} \begin{pmatrix} 1 & 1 \end{pmatrix}$, $ \textbf{k}_3=  k_{lat} \begin{pmatrix} 0 &  1 \end{pmatrix}$ , and $ \textbf{k}_4= k_{lat}/\sqrt{2} \begin{pmatrix} 1 & -1 \end{pmatrix} $, while $V_0$ is a parameter that regulates the strength of the potential.
This potential exhibits eightfold rotational symmetry, and it takes values between $0$ and $4V_0$, with a global maximum at $\textbf{r}=0$, which coincides with the center of the trap, see \figref[a-b]{geometry}. 
Its quasiperiodic nature arises from the superposition of wave vectors at angles of $\pi/4$, which causes the components of their wave vectors to be incommensurate.
We express lengths in units of $l_{osc} = \sqrt{\hbar/m\omega}$, and energies in units of the recoil energy $E_r = \hbar^2k_{lat}^2/2m$. Temperatures are scaled to the superfluid critical temperature of the non-interacting trapped boson gas, $ k_B T_c = \hbar \omega \sqrt{\frac{6}{\pi^2} N}$ \cite{Bagnato1991}.

We model interactions via a hard-core potential of scattering length $a_{2D}$. In two spatial dimensions, physical effects of interactions can be observed for exponentially small values of the scattering length \cite{pet00}. For this reason, we introduce the dimensionless parameter 
\begin{equation}
\tilde{g} = 2\pi \left( \ln \frac{l_{osc}}{a_{2D}}\right)^{-1}\,,
\end{equation}
which is related to the 2D mean-field parameter $g$ by $g = \hbar^2/m \tilde{g}$ when $\tilde{g} \ll 1$. In the same limit, it coincides with the effective interaction parameter used for trapped ultracold atoms in the quasi-two-dimensional regime, which is usually given as $ \tilde{g} = \sqrt{8\pi} l_z / a_{3D} $, $a_{3D}$ being the three-dimensional scattering length of the gas, and $l_z$ the trapping along the $z$ axis. Both quantities play no role in our model, which is purely two-dimensional, but the parameter $\tilde{g}$ serves as a bridge between the two approaches.

\begin{figure}[t!]
\includegraphics[width=\linewidth]{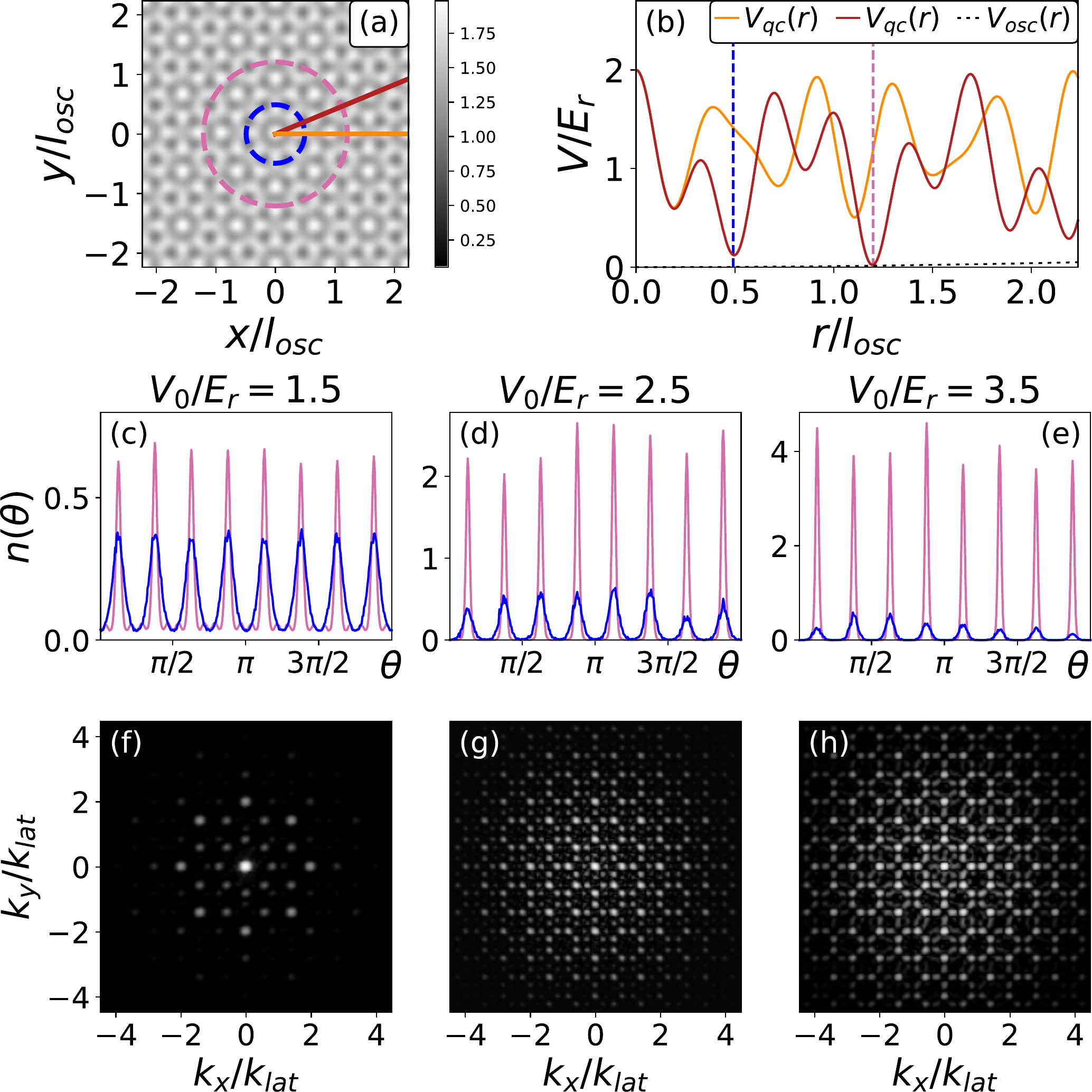}
\caption{\textit{Geometric features.} 
(a) 2D plot of the quasiperiodic potential $V_{qc}(\textbf{r})/E_r$ for $V_0=0.5\, E_r$. White regions correspond to peaks and shaded ones to wells. 
Two directions are highlighted, corresponding to $\theta=0$ (orange) and $\theta=\pi/8$ (brown). The deepest local minima lie on the $\theta=\pi/8$ line. Two circles are highlighted, crossing the eight local minima closer to the trap center (blue) and farther away (magenta). These sixteen sites are the most relevant to localization properties at the chosen value of the interaction. (b) $V_{qc}(r)$ is plotted along the two directions highlighted in (a), again at a value of $V_0 = 0.5\, E_r$. The vertical lines mark the deepest local minima, corresponding to the circles highlighted in (a). The dotted black line line represents the harmonic potential. (c-e) Plots of the boson density as a function of the angle, along the circles highlighted in (a), at $T/T_c = 0.25$. Different pictures correspond to different choices of the potential, $V_0 = 1.5\, E_r$ (left), $V_0 = 2.5\, E_r$ (middle), and $V_0 = 3.5 E_r$ (right). (f-h) Diffraction patterns, normalized to the peak density and in a logarithmic scale. The respective values of $V_0$ are the same as in the density profiles above.}
\label{fig:geometry}
\end{figure}
In our simulations we set the trap in such a way as to obtain a trap-center density close to experimental values. 
To draw a phase diagram in \figref[d]{phasediagram}, we choose a specific value of $\tilde{g} = 0.0217$ within reach of current experimental setups with quasi-$2$D Bose gases (see e.g. \cite{cho13}).

\myparagraph{Simulation methods and estimators} We make use of continuous-space Path Integral Monte Carlo (PIMC) for a number of particles up to $N=500$. The PIMC method can provide exact estimates of thermodynamic observables for quantum systems at finite temperature \cite{cep95, krauth2006statistical, PhysRevLett.96.070601, Boninsegni2006}. Each quantum particle is mapped into a classical polymer, and observables are sampled in the classical system. Polymers can then connect to each other, representing coherence and the emergence of superfluidity. 
In \figref[a-c]{phasediagram}, we show three PIMC shapshots of the superfluid, the BG, and the insulating phase, respectively. Corresponding angular densities, along the circles in \figref[a]{geometry}, are reported in \figref[c-d]{geometry}, while diffraction patterns are shown in \figref[f-h]{geometry} (a description of the estimators used, as well as some additional diffraction patterns at smaller $V_0$, can be found in \cite{SM}).

The hard-core interaction is implemented through the pair-product approximation \cite{bar79, cep95, pil06}, requiring, in two dimensions, the use of tables for the propagator. 

\begin{figure}[t!]
\includegraphics[width=\linewidth]{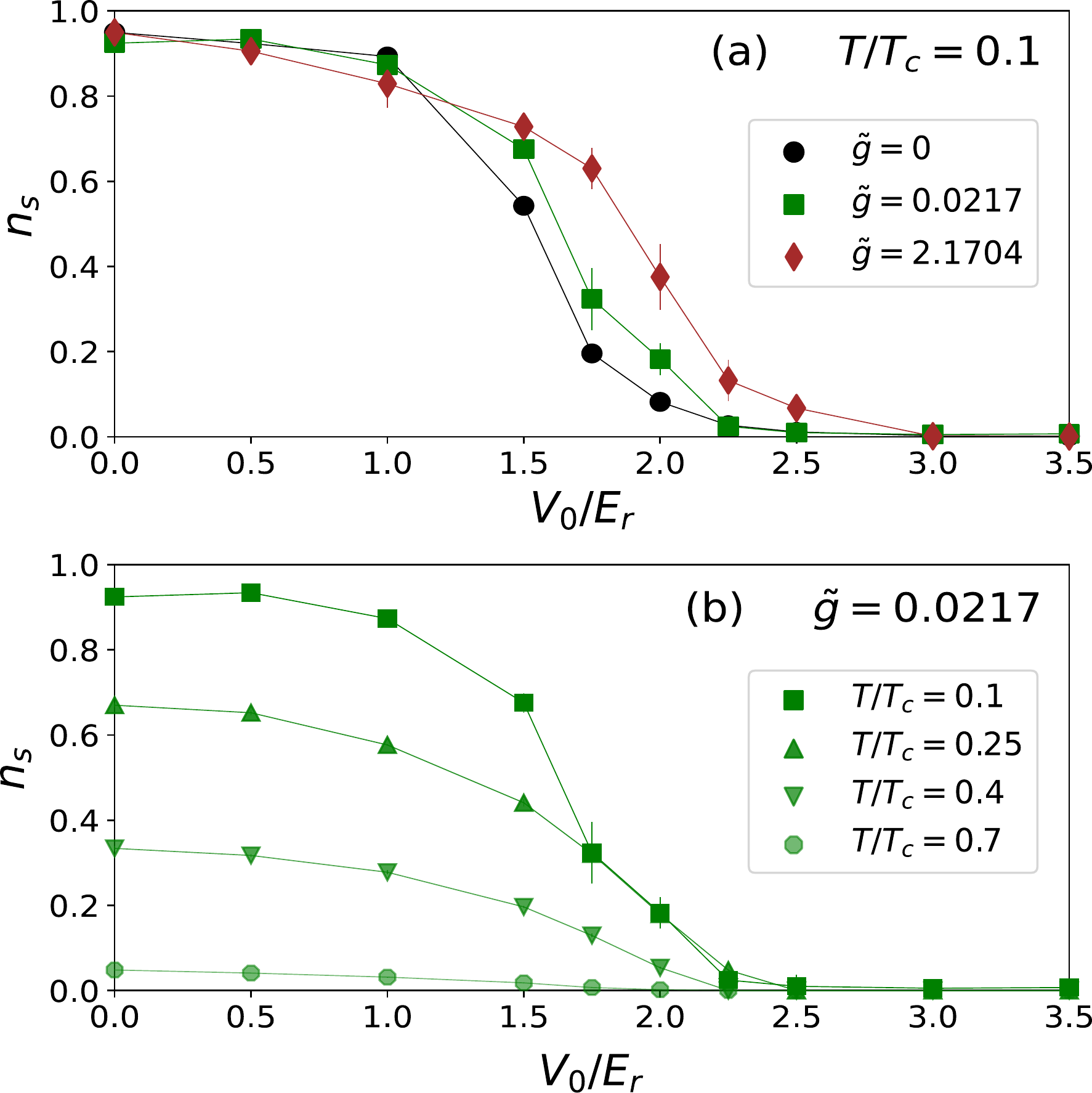}
\caption{\textit{Global superfluid fraction.} (a) $n_s$ as a function of the potential strength $V_0$, at fixed temperature $T/T_c=0.1$. Lines are guides for the eye.
The three sets of points are at interaction $\tilde{g} =0$ (black circles), $\tilde{g} =0.0217$ (green squares), and $\tilde{g} = 2.1704$ (brown diamonds). (b) $n_s$ against the potential parameter $V_0$, at fixed interaction parameter $\tilde{g}=0.0217$. Different sets of points correspond to different temperatures, $T/T_c= 0.1$ (squares), $T/T_c= 0.25$ (upward triangles), $T/T_c= 0.4$ (downward triangles), $T/T_c= 0.7$ (octagons). The points at $T/T_c=0.1$ are also displayed in (a) as green squares.}
\label{fig:globalns}
\end{figure}

In systems with periodic boundary conditions, the superfluid fraction is characterized by the well-known winding number estimator \cite{cep86}, which is not applicable to a trapped system. Instead, we employ the area estimator \cite{cep95,PhysRevB.84.014534}, which is directly related to the reduction of the moment of inertia associated with the emergence of superfluidity. The estimator is derived in its entirety in \cite{zen14}, and can be written as
\begin{equation} \label{global}
n_s = \frac{4 m^2}{\hbar^2 \beta} \frac{\langle A^2 \rangle - \langle A \rangle^2}{I_{cl}},
\end{equation}
where $A$ is the total area enclosed by the polymers. Customarily, the $\langle A \rangle^2$ term is neglected on the grounds of temporal invariance of the system dynamics. In the localized phase, as a symptom of ergodicity breaking, this term does not necessarily average to 0; it must then be kept into account, to give a meaningful estimate of the superfluid fraction.

Due to the presence of a harmonic trap, observables such as superfluid fraction and compressibility stop being homogeneous across the system. In order to investigate their behavior, local estimators have been introduced; one example is found in \cite{wes04}, where a local superfluid density and a local compressibility are used on-lattice to characterize a trapped Bose-Hubbard model. The extension of these local observables to the continuous case presents technical difficulties due to the noisy character of the estimators.
%and, at least in the case of the superfluid density, to the dependence on the local moment of inertia, which vanishes as $r\to 0$, causing large fluctuations close to the center of the trap.

Instead, we have chosen to focus on \textit{zonal estimators}, which aim at approximating the behavior of physical observables in finite portions of the system. We separate the simulation space into three regions, as depicted in \figref[a]{zonalplots}. The choice is made based on the arrangement of the sixteen central sites, which are the most relevant for localization. 
In each of the three regions, we measure a zonal compressibility, \begin{equation}\label{zonalcomp}
\kappa^{(z)} = \beta(\langle N^{(z)2} \rangle -\langle N^{(z)} \rangle^2),
\end{equation} and a zonal superfluid fraction, $n_s^{(z)}$, which is obtained by integration of the local estimator. A detailed discussion of the latter can be found in \cite{SM}.

\myparagraph{Global superfluidity} 
In \figref[a]{globalns}, we show the results for the global superfluid fraction at different values of $\tilde{g}$, at 
$T/T_c = 0.1$. 
We find that stronger interactions tend to increase the superfluid fraction at a given value of $V_0$, while also increasing the localization potential to higher values. At low temperature, this observable acts as a signature of the localization transition \cite{Sbroscia2020}. We observe that the presence of a weak harmonic potential does not significantly alter the localizing behavior with respect to the homogeneous case.

The question, then, is whether the transition persists at higher temperatures, when thermal fluctuations are not negligible, see \figref[b]{globalns}. 
As the temperature rises, in the absence of the quasiperiodic potential, the superfluid fraction decreases. The same behavior is visible for low strengths of the quasiperiodic potential. Distinctions between different temperatures appear as we move to larger values of $V_0$. At $T/T_c = 0.25$, the reduction of superfluidity by confinement is still essentially the same, indicating that ground-state physics is still dominant in the localization process. 
As we approach $T_c$, the superfluid signal reduces significantly; we will argue in the next section that this coincides with the reduction of superfluidity in the inner regions of the trap, and with the disappearance of the glass phase.

\myparagraph{Zonal estimators}
In a homogeneous system, in the grand canonical ensemble, it is possible to directly measure the compressibility; at the same time, the global superfluid fraction can be accessed through the winding number estimator, exploiting the presence of periodic boundary conditions. Together, these two quantities allow us to discriminate between the BG and the MI phases. With this method, it is not possible to directly characterize superfluid puddles; the reason being that the winding number estimator relies on particle paths crossing the whole system, so that a bounded superfluid region produces no signal. 
Conversely, in a finite system, superfluidity is related to the response to an applied angular velocity rather than to a linear velocity, the very principle that the area estimator is based on. If a region of local superfluidity is rotationally symmetric around the center of the trap, it is then possible to measure a finite superfluid response locally, even when its contribution to the global superfluid response is negligible. The use of zonal estimators enables us, to a certain extent, to identify superfluid puddles. Crucially, in the chosen geometry, one such puddle is expected to appear in the central region of the trap.

Zonal compressibility, as defined in eq.\eqref{zonalcomp}, measures particle fluctuations in different regions. Since we are working in the canonical ensemble, fluctuations are only due to translations and not to the creation and destruction of particles. The zonal estimator, then, effectively acts as a measure of particle localization in each region.
The discrimination between different phases proceeds as follows. In the SF phase, the superfluid fraction and the zonal compressibility are finite in all regions; bosons are superfluid in the whole trap, and they are able to move freely across it. In the insulating phase, on the other hand, both estimators are zero in all regions, as particles become fully localized and superfluidity is depleted. The BG phase presents a strongly suppressed compressibility, indicating that particles are unable to move between different regions, but the zonal superfluid fraction remains larger than zero in the central region.

\begin{figure}[t]
\includegraphics[width=\linewidth]{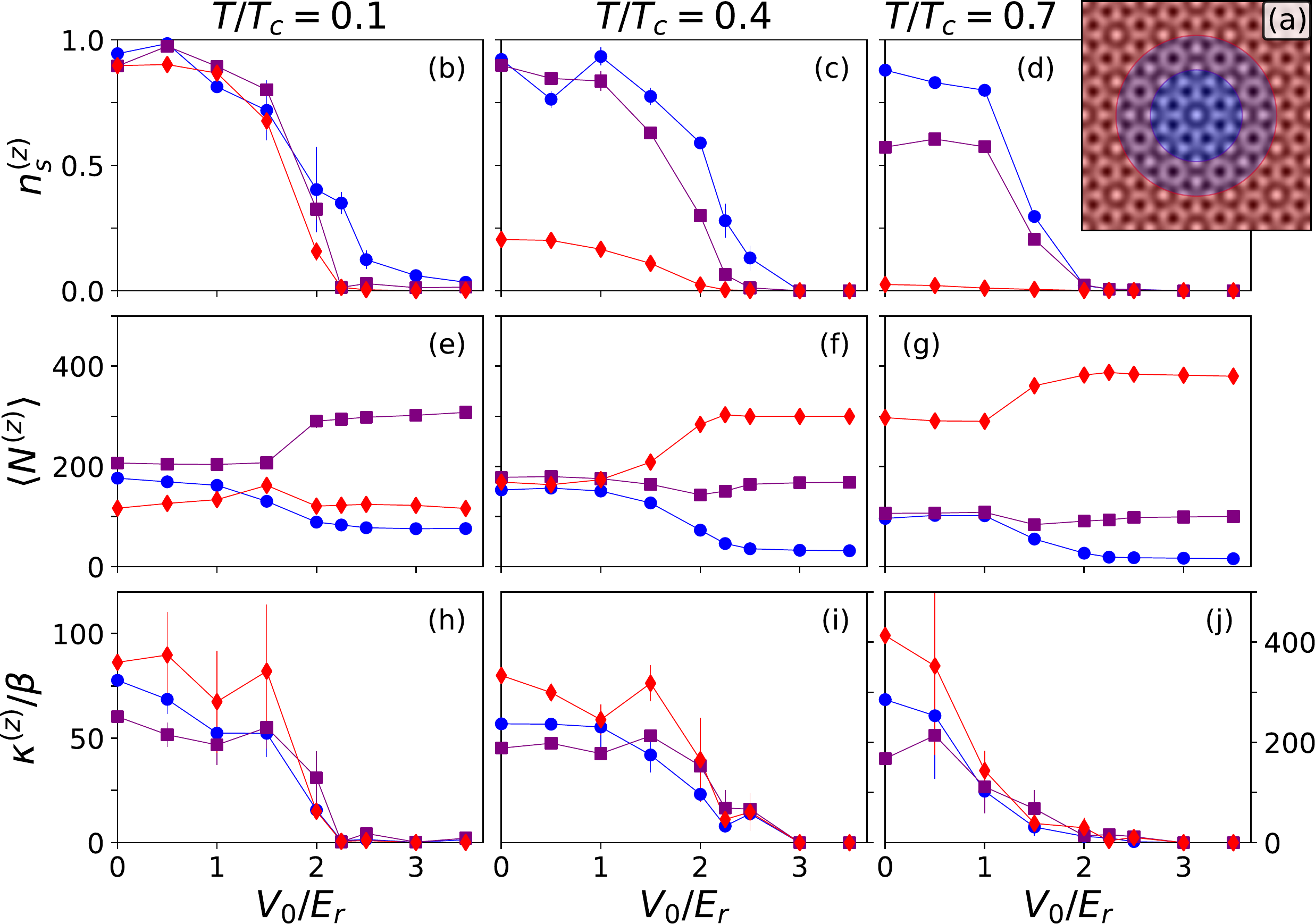}
\caption{\textit{Zonal estimators.} (a) 2D plot of the quasiperiodic potential; different regions correspond to $r<r_a$ (blue), $r_a < r < r_b$ (purple), and $r > r_b$ (red). (b-j) Zonal quantities at $\tilde{g}=0.0217$, at varying temperature and potential. Points mark simulation results. Lines are guides for the eye. Plots (b-d) display the zonal superfluid fraction $n_{s}^{(z)}$; (e-g) display the average number of particles in each region, $\langle N^{(z)} \rangle$; (h-j) display the zonal compressibility. Note that $\kappa^{(z)}/\beta = \langle N^{(z)2}\rangle - \langle N^{(z)} \rangle^2$. The scale of $\kappa^{(z)}/\beta$ in (h) is marked on the left, while (i) and (j) share the same scale, marked on the right. In all plots, the three sets of points correspond to the three regions depicted in (a):  $r<r_a$ (blue circles), $r_a < r < r_b$ (purple squares), and $r > r_b$ (red diamonds).}
\label{fig:zonalplots}
\end{figure}

Our results for $\tilde{g} = 0.0217$ are reported in \figref[b-j]{zonalplots}. Up to $T/T_c = 0.4$, the reduction in compressibility (third line) happens at the same values of $V_0$ as the depletion of the global superfluidity, and of the zonal superfluid fraction in the outer regions. The superfluid fraction in the inner region, on the contrary, remains small but distinctly larger than zero, signaling the presence of the BG, up to some higher value of $V_0$. The information thus obtained is used to draw the ``phase diagram'' in \figref[d]{phasediagram}. The configurations of \figref[a-c]{phasediagram} represent snapshots of the particle paths at a given simulation step, with connected particle paths giving an indication of coherence. It is again possible to distinguish between a SF phase, where coherence is established among a large number of particles; an insulating phase that exhibits full localization in lattice sites; and the BG, which displays coherence only in the central region. 

At finite $T$, it is known that depletion of the superfluid begins from the edges and proceeds to the center of the trap, so that, close to $T_c$, only the bosons in the central region are superfluid. This behavior is well characterized by the zonal superfluid fraction: while, at $T/T_c=0.1$, all three regions are equally superfluid, as $T$ increases we see that superfluidity is depleted starting from the outer region (green lines). Nonetheless, we have chosen to label these points as ``superfluid'' in \figref[d]{phasediagram}. 

\myparagraph{Discussion and conclusions}
We employed PIMC simulations at finite temperature to determine the ``phase diagram'' of 2D trapped bosons in quasiperiodic potentials.
We point out that, as shown in \figref[b]{geometry}, the harmonic potential is much weaker than the quasiperiodic one; while it enforces a circular symmetry on the system, and it selects a certain region of space, it has no impact on the actual distribution of the bosons in the minima within this region. In this respect, our results could be compared with those obtained in homogeneous systems of similar spatial extensions.
We found a superfluid and an insulating phase, as well as a normal fluid at high temperatures.
At intermediate strengths of the quasicrystalline potential, the system exhibits a BG phase. The values of densities and interaction strengths chosen are comparable with those used in state-of-the-art experimental setups with ultracold atoms. Notably, the BG is stable up to relatively high temperatures $T/T_c \simeq 0.4$.

A physical implementation of this proposal can be realized using $^{23}$Na as done in \cite{cho13}, where values of $\tilde{g} \sim 0.01$ have been reached with a longitudinal trapping frequency $\omega_z = 2\pi \times 370$ Hz, leading to $l_z \approx 840$ nm.
Setting $\tilde{g} = 0.0217$ as in the simulations of \figref{zonalplots}, with the same harmonic confinement, we get $a_{3D} \approx 70\; a_0$.
Alternatively, $^{39}$K can be employed \cite{Sbroscia2020}. 
Concretely, using $\lambda_{lat} = 725$ nm, $l_{osc} \simeq 1.15\, \mu$m, and setting $\tilde{g} = 0.0217$ as above, in the Thomas-Fermi limit, this leads to the center-trap density $n(0) \simeq  0.68\times 10^{14}$ m$^{-2}$,
comparable to peak density in the experiment, where $n_\text{exp} \simeq 1.24 \times 10^{14}$ m$^{-2}$ \cite{Sbroscia2020}. 

Regarding the access to zonal quantities in real platforms, we expect that single-site resolution in lattice geometries should allow to extract local particle number fluctuations, to measure the zonal compressibility. The measurement of the zonal superfluid fraction is clearly more challenging. At the present time, global superfluidity has been measured in a very limited number of experimental setups \cite{guo20, der16}.

In conclusion, our work offers a strong motivation for further investigation of interacting quasicrystalline phases in current ultracold atom platforms, as well as a benchmark for future studies into the thermodynamics and dynamics of systems in quasiperiodic potentials at finite temperature. Further exploration of quasicrystalline properties induced by an external potential will proceed in parallel with the study of excitation spectra \cite{PhysRevLett.123.223201} and exotic self-assembled quantum many-body phases with non-local interaction potentials \cite{Barkan14, pup20, abr20,PhysRevA.96.013627,PhysRevLett.119.215302}.

\begin{acknowledgments}
\myparagraph{Acknowledgments}
We thank the High Performance Computing Center (NPAD) at UFRN and the Center for High Performance Computing (CHPC) in Cape Town for providing computational resources.
\end{acknowledgments}

%%%%%%%%%%%%%%%%%%%%%%%%%%%%%%%%%%%%%%%
%%%%%%%%%%%%%%%%%%%%%%%%%%%%%%%%%%%%%%%
%%%%%%%%%%%%%%%%%%%%%%%%%%%%%%%%%%%%%%%
%\bibliographystyle{apsrev4-2}[longbibliography]
%apsrev4-2.bst 2019-01-14 (MD) hand-edited version of apsrev4-1.bst
%Control: key (0)
%Control: author (8) initials jnrlst
%Control: editor formatted (1) identically to author
%Control: production of article title (0) allowed
%Control: page (0) single
%Control: year (1) truncated
%Control: production of eprint (0) enabled
%

\clearpage

%\phantomsection
\renewcommand{\theequation}{S.\arabic{equation}}
\setcounter{equation}{0}
\renewcommand{\thefigure}{S.\arabic{figure}}
\setcounter{figure}{0}

\section{Supplemental material}

\begin{figure*}[t]
\includegraphics[width=\textwidth]{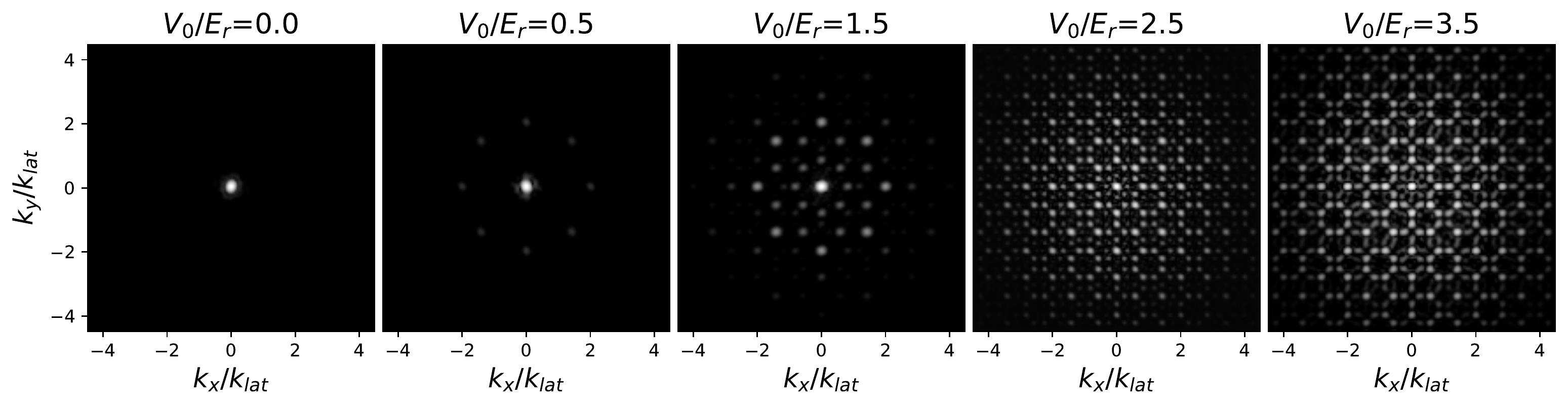}
\caption{Diffraction patterns from the bosons at $T/T_c=0.25$ and $\tilde{g} = 0.0217$. The intensity is normalized to 1 and in log scale.}
\label{fig:sup_diffs}
\end{figure*}

\myparagraph{Superfluidity and Bose-Einstein Condensation in two dimensions}

In this section, we clarify some confusion that may arise regarding the significance of superfluidity in our system, and its relation to the presence, or lack, of a Bose-Einstein condensate (BEC).

It is well-known that, in a homogeneous continuous system in $d=2$, there can be no second order phase transition, as fluctuations prevent long-range order from being established \cite{Hohenberg1967, Mermin1966}. In the case of Bose-Einstein condensation, the order parameter is represented by the condensate fraction \cite{pitaevski2003bose}; therefore, there can be no BEC in a homogeneous system in 2D.

On the other hand, the superfluid fraction is not an order parameter, and can instead  be characterized as a response function to an external velocity field, a property that has been extensively used to characterize superfluidity through a reduction of the moment of inertia (we also do so in the next section). In this sense, it can be different from zero even when long-range order is absent. In $d=2$, even when long-range order is forbidden, a different kind of quasi-long-range order can be formed in the context of the Berezinskii-Kosterlitz-Thouless (BKT) transition, which leads to a non-zero superfluid fraction below a certain temperature \cite{KosterlitzThouless1973, ber70, cha95}.

When harmonic trapping is introduced, the system is not homogeneous anymore, and it is possible for the system to display BEC in 2D; this is indeed the case for non-interacting bosons in $d=2$. The question of whether interacting bosons in a trap undergo a transition of the BEC or BKT kind has led to investigations of what is called the BEC/BKT crossover, both theoretically and experimentally \cite{hol07, fle15}.

In this paper, we take the critical temperature of the $d=2$ trapped Bose gas as a reference point, but we do not concern ourselves with the intricacies related to boson condensation and the BEC/BKT crossover. For our purposes, what is important is that we can distinguish superfluid and insulating phases, and our methods, as described in the next section, rely only on the definition of superfluidity as a response function, with no explicit reference to condensation.

\myparagraph{Details on the Path integral Monte Carlo method}  The core of the method lies in the application of Feynman's path integral to the partition function of a quantum system at finite temperature \cite{fey98, fey10}. Thermodynamic properties can then be measured on an equivalent, classical system, where each quantum particle is represented by a classical polymers. Quantum concepts, such as coherence and superfluidity, can be mapped across the equivalence as properties of the polymers, and can consequently be sampled by employing Monte Carlo procedures such as the Metropolis algorithm. In addition, we use the canonical Worm algorithm \cite{PhysRevLett.96.070601, Boninsegni2006} to efficiently sample configurations of connected polymers, which are crucial to the understanding of superfluidity. Reviews of and introductions to PIMC can be found in \cite{cep95, krauth2006statistical}.

The advantage of path integral techniques lies in their ability to determine the thermodynamic properties of the system starting from its basic constituents - the atoms and the microscopic interaction - within a precision limited only by numerical and statistical errors. In practice, the equivalence is realized approximately by breaking up the imaginary time interval $\beta$ into smaller intervals $\tau = \beta /M$.  To each particle $i$ corresponds, then, a classical polymer made of $j=1\dots M$ beads, connected with each other through harmonic springs. Errors introduced by the equivalence are reduced as $M$ increases. 

The basic version of our algorithm makes use of the harmonic propagator to efficiently simulate the behavior of bosons in the trapping potential, while the lattice is taken into account as en external potential in the sampling rates. The hard-core interaction is implemented through the pair-product approximation, requiring, in two dimensions, the use of tables for the propagator \cite{bar79, cep95, pil06}.

\myparagraph{Zonal superfluid fraction}

We define the zonal superfluid estimator, which was referenced in the main text. We begin with a quick review of the local estimator \cite{kwo06}.
	
In the context of the two-fluid model \cite{tis38, lon54}, the onset of superfluidity is described by separating the fluid into two components, a superfluid of density $\rho_s$ and a normal one of density $\rho_n$, contributing to the total density of the fluid:
\begin{equation}
\rho = \rho_n + \rho_s,    
\end{equation} 
The ratio of the superfluid density to the total one is the superfluid fraction,
\begin{equation}
n_s = \frac{\rho_s}{\rho}.    
\end{equation}

The two components have different properties in terms of flow and entropy transport; in particular, the superfluid component displays zero viscosity, and is therefore unresponsive to the application of external velocity fields. When we consider angular velocities, this leads to a reduction of the total moment of inertia, compared to a classical fluid in the same conditions. This relationship is stated as
\begin{equation}
n_s = 1 - \frac{I}{I_{cl}},
\end{equation}
where $I$ is the measured moment of inertia, which only the normal component contributes to, while $I_{cl}$ is the classical moment of inertia, which is the one the same mass of fluid would have if it behaved classically.

In the context of PIMC, the expectation value of the angular momentum is given in terms of the area encircled by particle paths, leading to the estimator
\begin{equation} \label{sm_globsl}
n_s = \frac{2 m}{\lambda \beta} \frac{\langle A_z^2 \rangle}{I_{cl}},
\end{equation}
which is equation \eqref{global} in the main text, where we omitted the non-ergodic term for brevity, and $\lambda = \hbar^2/2m$. In this expression,
\begin{equation}
A_z =  \frac{1}{2} \sum_{i=1}^{N} \sum_{j=1}^{M} \textbf{r}_{i,j} \times \textbf{r}_{i,j+1}
\end{equation}
is the total area enclosed by particle paths, and $\textbf{r}_{i,j}$ is the position of the $j$-th bead in the $i$-th particle.

We can manipulate the equations above to give
\begin{equation}\label{eq_iicl}
I = I_{cl} (1 - n_s) = I_{cl} - \frac{2 m}{\lambda \beta} \langle A_z^2 \rangle.    
\end{equation}

In inhomogeneous systems, the fields describing the two components acquire a spatial dependence, and so does the superfluid fraction itself:
\begin{equation}
\rho(\textbf{r}) = \rho_n(\textbf{r}) + \rho_s(\textbf{r}),    
\end{equation} 
\begin{equation}
n_s(\textbf{r}) = \frac{\rho_s(\textbf{r})}{\rho(\textbf{r})}.    
\end{equation}

This local superluid fraction can be characterized by breaking up the estimator \eqref{sm_globsl} into local contributions.

$I_{cl}$ is written explicitly as
\begin{equation}
I_{cl} = \int d\textbf{r} \; \rho(\textbf{r}) r^2.   
\end{equation}
Conversely, the measured moment of inertia is calculated by considering only the contribution of the normal component:
\begin{equation}
I = \int d\textbf{r} \; \rho_n(\textbf{r}) r^2 = \int d\textbf{r} \; \left[\rho(\textbf{r}) - \rho_s(\textbf{r})\right] r^2 = I_{cl} - \int d\textbf{r} \; \rho_s(\textbf{r}) r^2.
\end{equation}
This, by comparison with \eqref{eq_iicl}, leads us to
\begin{equation} \label{eq_nsicl}
\int d\textbf{r} \; \rho_s(\textbf{r}) r^2 = n_s I_{cl} = \frac{2 m}{\lambda \beta} \langle A_z^2 \rangle.
\end{equation}
    
A possible definition then suggests itself, as
\begin{equation}
	\rho_s(\textbf{r}) = \frac{2 m}{\lambda \beta}\frac{ \langle A_z A_z(\textbf{r}) \rangle }{r^2};
\end{equation}
this will integrate to the appropriate amount as long as $ \int d\textbf{r} \; A_z(\textbf{r}) = A_z $. The most common choice \cite{kwo06} is to define
\begin{equation}
	A_z(\textbf{r}) =  \frac{1}{2} \sum_{i=1}^{N} \sum_{j=1}^{M} \textbf{r} \times \textbf{r}_{i,j+1} \delta(\textbf{r}-\textbf{r}_{i,j}). 
\end{equation}

The $r^2$ term in the denominator is sometimes named the ``local contribution to the classical moment of inertia''. This is not entirely correct, since the local contribution is actually $\rho(\textbf{r}) r^2$. It is possible to express the decomposition of the superfluid fraction so that the local moment of inertia becomes directly relevant. From \eqref{eq_nsicl}, we find that
\begin{equation}
n_s = \frac{1}{I_{cl}} \int d\textbf{r} \; \rho_s(\textbf{r}) r^2 = \frac{1}{I_{cl}} \int d\textbf{r} \; n_s(\textbf{r}) \rho(\textbf{r}) r^2
\end{equation}
meaning that the global superfluid fraction is given by the average of the local superfluid fraction, weighted by the local moment of inertia.

As we mentioned in the main text, the local estimator is noisy and difficult to sample, especially in the localized phase. We can, however, exploit the integral decomposition to define superfluid fractions in different regions of the system. Given a region $A$, we can write 
\begin{equation}
n^A_s = \frac{1}{I^A_{cl}} \int_A d\textbf{r} \; \rho_s(\textbf{r}) r^2 = \frac{1}{I^A_{cl}} \int_A d\textbf{r} \; n_s(\textbf{r}) \rho(\textbf{r}) r^2,
\end{equation}
with the same definitions as before, but limiting the integration to the $A$ region. If the system is partitioned in a finite number of regions $A$, $B$..., we can then recover the global superfluid fraction as 

\begin{equation}
n_s = \frac{I^A_{cl}}{I_{cl}} n^A_s + \frac{I^B_{cl}}{I_{cl}} n^B_s + \dots
\end{equation}

This is, again, an average of the superfluid fractions of each region, weighted by the respective moment of inertia. Crucially, this decomposition shows that a region can have a finite superfluid fraction, but still give a negligible contribution to the global $n_s$, if the associated moment of inertia is small. This is the case for regions close to the trap center.

\myparagraph{Density profiles}

In the context of PIMC, two main ways are available to display the spatial configuration of the system. 
The first is to select a system configuration at a given simulation step, and to plot the position of each bead $\textbf{r}_i^j$, drawing a line between each pair of connected beads. The resulting figures are usually called snapshots. One advantage of this approach is that it allows to explicitly display connections between different particles, and therefore to obtain a visual representation of coherence. Such snapshots are the ones that we plot in \figref[a-c]{phasediagram}.
The second method is to plot density profiles, which are obtained as averages over simulation steps, as well as over the positions of all different beads associated to each particle. In continuous space, the average is usually performed  by separating the simulation area into bins, and counting the number of beads in each at every simulation step. To obtain the density profiles shown in \figref[c-e]{geometry}, we counted particles in 360 bins in correspondence of the circles drawn in \figref[a]{geometry}.

\myparagraph{Diffraction patterns}

The structure factor is a quantity directly related to diffraction patterns, that can be observed experimentally in scattering experiments. It is defined, for a particle density $n(\textbf{r}) = \sum_i \delta(\textbf{r}_i)$, as
\begin{equation}
	\label{struct}
	I(\textbf{q}) = \langle n(\textbf{q}) n(-\textbf{q}) \rangle, 
\end{equation}
with 
\begin{equation} 
n(\textbf{q}) = \int d^2\textbf{r} \; e^{-i\textbf{q}\cdot\textbf{r}} n(\textbf{r}) = \sum_{j} e^{-i\textbf{q}\cdot\textbf{r}^j} 
\end{equation}
the Fourier transform of the particle distribution \cite{cha95}. To measure this quantity, we compute the sum and average over beads and simulations steps, similarly to what we do for the density profiles. This is done for a set of wavevectors, on the vertices of a grid in $\textbf{q}$-space.

\begin{figure}[t!]
\includegraphics[width=\linewidth]{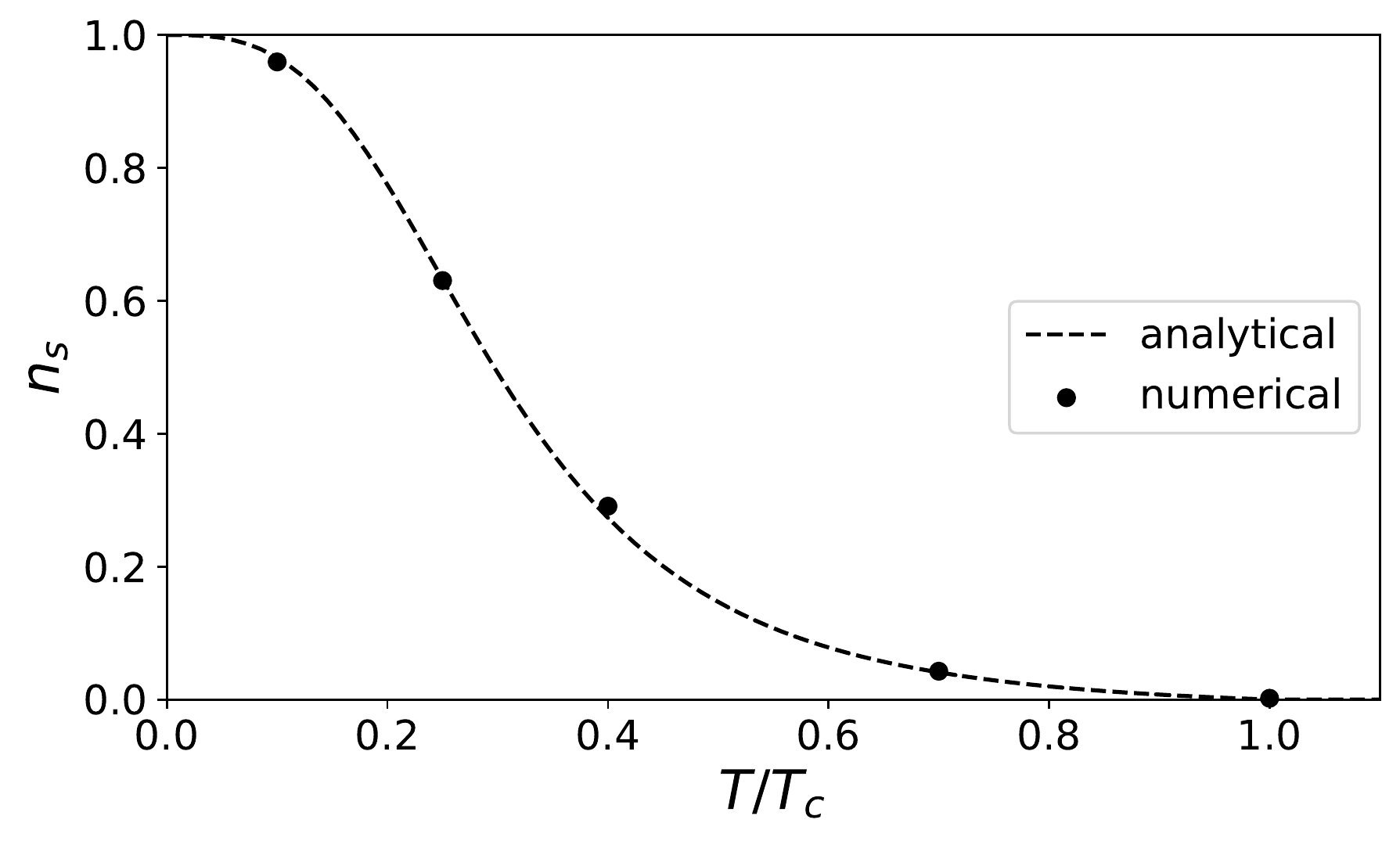}
\caption{Depletion of the global superfluid fraction at different values of $T$, in the non-interacting case. The dashed line is obtained analytically, while the dots are simulation results.}
\label{fig:sup_temps1}
\end{figure}

In \figref{sup_temps1}, we display some diffraction patterns. These are the same reported in \figref[f-h]{geometry} of the main text, with the addition of the values of $V_0=0$ and $V_0/E_r=0.5$. As we could expect, the structure factor evolves from a single peak in the fluid phase to a typical quasicrystalline pattern.

\begin{figure}[b!]
\includegraphics[width=\linewidth]{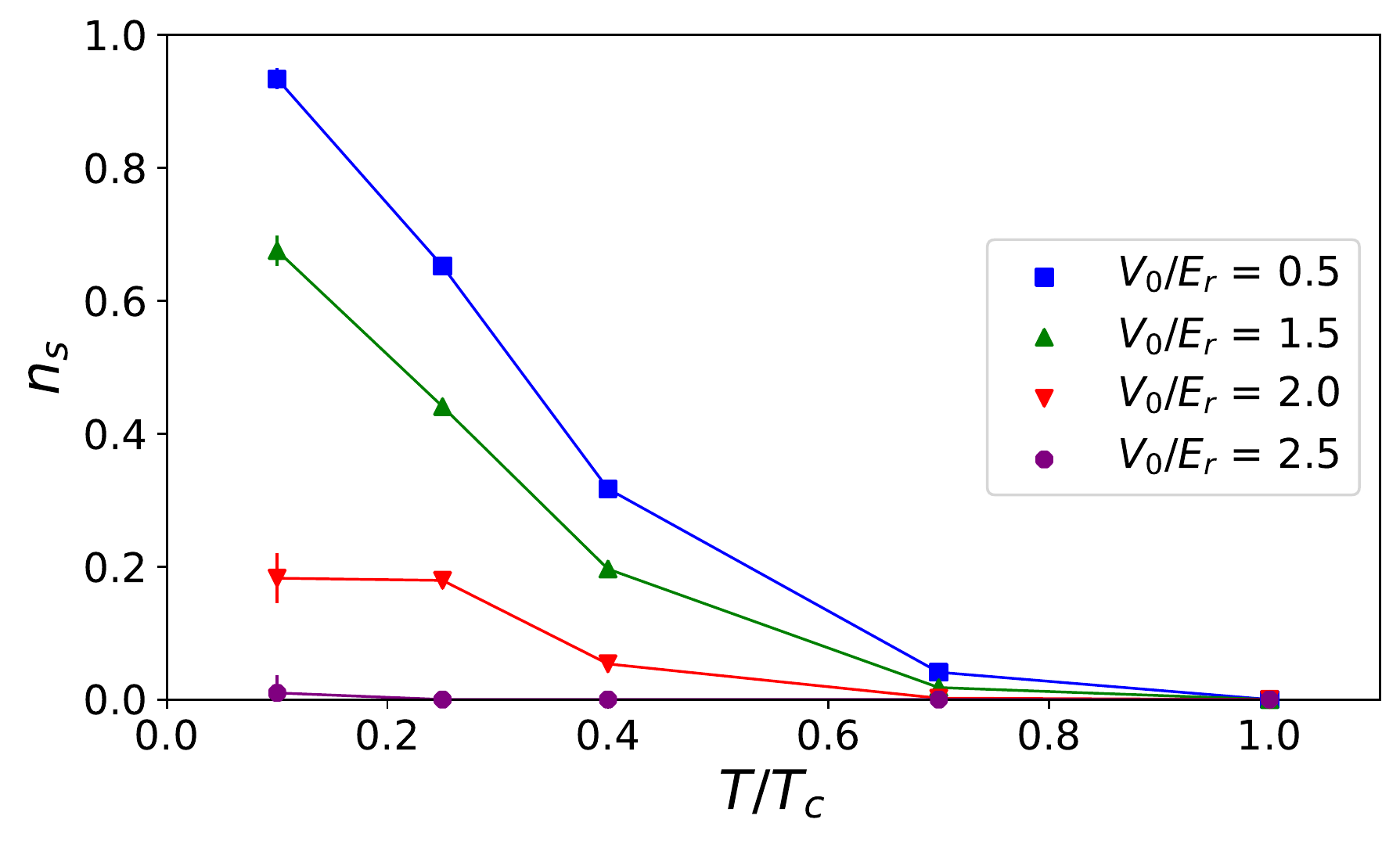}
\caption{Global superfluid fraction at different values of $T$, at $\tilde{g}=0.0217$. The lines are a guide for the eye.}
\label{fig:sup_temps2}
\end{figure}

\myparagraph{Temperature behavior of the non-interacting gas}

For free bosons in a harmonic trap, the temperature behavior of the global superfluid fraction can be predicted by analytical estimates. First, $n_s$ is related to the number of particles in the condensate, from considerations on its moment of inertia \cite{sch00}: 
\begin{equation} \label{sm_2dsuper}
n_s(T) \simeq \frac{1}{1 + \frac{N - \langle N_0 \rangle}{\langle N_0 \rangle} \frac{2k_BT}{\hbar\omega}} , 
\end{equation}
where $\langle N_0 \rangle$ is the number of particles in the condensate at temperature $T$. This quantity can be directly computed from the energy density of states, to give 
\begin{equation} \label{sm_condensation}
\langle N_0 \rangle = N - \int_0^{\infty} d\epsilon \rho(\epsilon) n(\epsilon) = N - \frac{k_B^2T^2}{\hbar^2\omega^2} \frac{\pi^2}{6},
\end{equation}
$\rho(\epsilon)$ being the energy density of states. Plugging \eqref{sm_condensation} into \eqref{sm_2dsuper}, we obtain a formula for the global superfluid fraction as a function of the temperature, which we plot as a dashed line in \figref{sup_temps1}. The dots are values of $n_s$ estimated from our simulations, which show perfect agreement with the analytical prediction.

In \figref[b]{globalns}, we showed plots of $n_s$ as a function of $V_0$, at different temperatures, for the interacting gas. In \figref{sup_temps2}, instead, we keep $V_0$ fixed and plot $n_s$ against $T$.

\end{document}